\newif\ifpdf
\ifx\undefined\pdfoutput \pdffalse \else \pdftrue \fi
\documentclass{elsart}
\usepackage{marvosym,changebar,dsfont,tabularx,graphicx,amsmath}

\ifpdf
   \DeclareGraphicsExtensions{.pdf}
\else
   \DeclareGraphicsExtensions{.eps}
\fi

\begin{document}
\begin{frontmatter}

\title{Generalised Recurrence Plot Analysis for Spatial Data}
\author{Norbert Marwan\corauthref{cor}},
\corauth[cor]{Corresponding author.}
\author{J\"urgen Kurths}
\address{Institute of Physics, University of Potsdam, 14415 Potsdam, Germany}
\author{Peter Saparin}
\address{Department of Biomaterials, Max Planck Institute of Colloids and Interfaces, 14424 Potsdam-Golm, Germany}

\begin{abstract}
Recurrence plot based methods are highly efficient and
widely accepted tools for the
investigation of time series or one-dimensional data.
We present an extension of the recurrence plots and their 
quantifications in order to study recurrent structures in higher-dimensional 
spatial data. The capability of this extension is illustrated on 
prototypical 2D models. Next, the tested and proved 
approach is applied to assess the bone 
structure from CT images of human proximal tibia. We find that 
the spatial structures in trabecular bone become more self-similar 
during the bone loss in osteoporosis. 
\end{abstract}
\end{frontmatter}

\section{Introduction}
Recurrence is a fundamental property of many dynamical
systems and, hence, of various processes
in nature. A system may strongly diverge, but after
some time it recurs ``infinitely many times as close as one 
wishes to its initial state'' \cite{poincare1890}.
The investigation of recurrence
reveals typical properties of the system and may help to 
predict its future behaviour. With the study of nonlinear
chaotic systems several methods for the investigation of
recurrences have been developed. The method
of recurrence plots (RPs) was introduced by Eckmann et~al.~\cite{eckmann87}.
Together with different RP quantification approaches
\cite{webber94,marwan2002herz}, this method has
attracted growing interest for both theory and applications
\cite{marwan2003diss}.

Recurrence plot based methods have been succesfully applied
to a wide class of data from physiology, geology,
physics, finances and others. They are especially suitable for
the investigation of rather short and nonstationary data.
This approach works with time series or phase space reconstructions
(trajectories), i.\,e.~with data which are 
at least one-dimensional. 

Recurrences are not restricted to one-dimensional time 
series or phase space trajectories. Spatio-temporal processes
can also exhibit typical recurrent structures.
However, RPs as introduced in \cite{eckmann87} cannot be directly applied to 
spatial (higher-dimensional) data. One possible way to study the
recurrences of spatial data is to separate the higher-dimensional
objects into a large number of one-dimensional data series, and to analyse 
them separately \cite{vasconcelos2006}. A more promising approach is
to extend the one-dimensional approach of the recurrence plots to
a higher-dimensional one.

In the presented work, we focus on the analysis of snapshots of spatio-temporal
processes, e.\,g., on static images. An extension of recurrence plots and
their quantification to higher-dimensional data is suggested.
This extension allows us to apply this method directly to 
spatial higher-dimensional data, and, in particular, 
to use it for 2D image analysis. We apply this method to 2D pQCT  
human bone images in order to investigate differences in 
trabecular bone structures at different stages of osteoporosis.

\section{Recurrence Plots}

The initial purpose of recurrence plots was the visualisation of
recurrences of system's states $\vec{x}_i$ in a phase space (with dimension $m$)
within a small deviation $\varepsilon$ \cite{eckmann87}. The RP
efficiently visualises recurrences even for high dimensional
systems. A recurrence of a state at time $i$ at a different
time $j$ is marked within a two-dimensional squared matrix
with ones and zeros dots (black and white points in the plot), 
where both axes represent time. The RP can be formally expressed by the matrix
\begin{equation}\label{eq_rp}
\mathbf{R}_{i,j} = \Theta\left(\varepsilon-\left\|\vec x_{i} - \vec x_{j}\right\|\right),
\quad \vec x_{i} \in \mathds{R}^m,\quad i,j=1 \ldots N,
\end{equation}
where $N$ is the number of considered states $\vec x_i$, 
$\varepsilon$ is a threshold distance (an arbitrary deviation range within 
a recurrence is defined), \mbox{$\Vert\cdot\Vert$} denotes
a norm and $\Theta (\cdot)$ is the Heaviside function. 

\begin{figure}[hbtp] 
\centering \includegraphics[width=0.3\columnwidth]{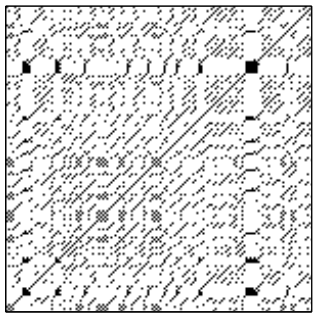} 
\caption{Example of a recurrence plot for the logistic map 
($x_{i+1} = ax_i (1 - x)$ with control parameter $a=3.9767$).
The RP consists of single dots and line structures.}\label{fig_rp_smallscalestructures}
\end{figure}

It should be emphasised that this method is a pairwise comparison
of system's states at different times along a phase 
space trajectory, which is -- although 
lying in an $m$-dimensional space -- a one-dimensional curve. The axes of the
RP correspond to the time which is given by pursueing a
state on the trajectory. Diagonal lines in an RP represent epochs of similar
dynamical evolution of the analysed system. For $i=j$ we get
the {\it line of identity} (LOI), $\mathbf{R}_{i,i} \equiv 1 \,|_{i=1}^N$, which is
the main diagonal line in the RP (Fig.~\ref{fig_rp_smallscalestructures}).

Instead of using the system's states $\vec{x}_i$ which are
often unknown, RPs can be created by only using a single time series or a
reconstruction of the phase space vectors (e.\,g.~by using
time-delay embedding, \cite{takens81}). Such applications 
to experimental data have expanded the utilisation of RPs from
a tool for the investigation of {\it deterministic phase space dynamics} to a tool
for the investigation of {\it similarity and transitions in data series}, 
without the rather strong requirement that the data must be from a deterministic
dynamical process.
The idea of such a similarity plot is not new and can be found
in publications earlier than \cite{eckmann87}, e.\,g.
in \cite{maizel1981}.
This alternative understanding was (unconsciously) the base of the
ever increasing amount of application of RPs in data analysis.
However, in its present state the RP technique could not
be applied on higher-dimensional spatial data.

The initial purpose of RPs was the visual inspection of 
the behaviour of phase space trajectories. The appearance of 
RPs gives hints about the characteristic time evolution 
of these trajectories 
\cite{marwan2003diss}. A closer inspection of 
RPs reveals small-scale structures which 
are {\it single dots}, {\it diagonal lines} as well as {\it vertical}
and {\it horizontal lines} (Fig.~\ref{fig_rp_smallscalestructures}). 

A {\it diagonal line} 
$\mathbf{R}_{i+k, j+k} \equiv 1\,|_{k=0}^{l-1}$
(where $l$ is the length of the diagonal line)
occurs when one segment of the trajectory runs parallel to another one,
i.\,e.~the trajectory re-visits the same region of the phase space at
different time intervals. 
The length of this diagonal line is determined by 
the duration of intervals with similar local behaviour 
of the trajectory segments. 
We define a line in the RP as a diagonal line of length $l$, if it fulfils the condition
\begin{equation}\label{eq_diagonalline}
 \left( 1 - \mathbf{R}_{i-1,j-1} \right) 
 \left( 1 - \mathbf{R}_{i+l,j+l} \right) 
\prod_{k=0}^{l-1} \mathbf{R}_{i+k,j+k}
     \equiv 1.
\end{equation}

A {\it vertical (horizontal) line} 
$\mathbf{R}_{i, j+k} \equiv1 \,|_{k=0}^{v-1}$
(where $v$ is the length of the vertical line)
marks a time interval in which a system's state does not change 
in time or changes very slowly. 
It looks like the state is trapped for some time, which is a typical
behaviour of laminar states \cite{marwan2002herz}. Because
RPs are symmetric about the LOI by definition (\ref{eq_rp}), each vertical
line has a corresponding horizontal line. Therefore,
only the vertical lines are henceforth considered. Combinations of
vertical and horizontal lines form rectangular clusters
in an RP.
We define a line as a vertical line of length $v$, if it fulfils the condition
\begin{equation}\label{eq_verticalline}
 \left( 1 - \mathbf{R}_{i,j-1} \right) 
 \left( 1 - \mathbf{R}_{i,j+v} \right) 
\prod_{k=0}^{v+1} \mathbf{R}_{i,j+k}
     \equiv 1.
\end{equation}

These small-scale structures are used for the quantitative 
analysis of RPs (known as recurrence quantification analysis, RQA). 
Using the distributions of the lengths
of diagonal lines $P(l)$ or vertical lines $P(v)$, different
measures of complexity have been introduced (cf.~\cite{marwan2003diss}
for a comprehensive review of definitions and descriptions of 
these measures). Here we generalise the measures
recurrence rate $RR$, determinism $DET$, averaged diagonal
line length $L$, laminarity $LAM$ and trapping time $TT$
in order to quantify higher-dimensional data.
(cf.~Tab.~\ref{eq_rqa}).

Several measures need a predefined minimal length $l_{min}$ or $v_{min}$,
respectively, for the definition of a diagonal or vertical line. These
minimal lengths should be as minimal as possible in order to cover 
as much variation of the lengths of these lines.
On the other hand, $l_{min}$ and $v_{min}$ should be large enough to exclude
line-like structures which represent only single, non-recurrent states, which may
occur if the threshold $\varepsilon$ is chosen too large or if the data 
have been smoothed too stronlgy before computing the RP.

RQA was successfully applied for example for
the detection of transitions in event related EEG potentials \cite{marwan2004},
the study of interrelations between El Ni\~no and climate in the past \cite{marwan2003climdyn},
the investigation of economic data series \cite{gilmore96},
of nonlinear processes in electronic devices \cite{elwakil99}
or the study of transitions in chemical reactions \cite{rustici99}.
For a number of further applications see, e.\,g., \cite{marwan2003diss} or 
www.recurrence-plot.tk.

\begin{table}
\caption{Generalised recurrence quantification measures 
for spatial data of dimension $d$ and with 
$\vec \imath,\vec \jmath \in \mathds{Z}^d$.
Note that these measures assess recurrence information
in terms of length while the original RQA measures
quantify it in terms of time.}\label{eq_rqa}
\begin{tabularx}{\textwidth}{XXX}
\hline
RQA measure&equation&meaning\\
\hline
\hline
    recurrence rate	&
    $RR =  \frac{1}{N^{2d}} \sum \limits_{\vec \imath,\vec \jmath}^N \mathbf{R}_{\vec \imath,\vec \jmath}$ &
    percentage of recurrent states in the system; probability of the recurrence
    of any state
\\
    determinism	&
    $ DET_{HS}=\frac{\sum_{l=l_{min}}^N l\, P(l)}{\sum_{\vec \imath,\vec \jmath}^N \mathbf{R}_{\vec \imath,\vec \jmath}}$ &
    percentage of recurrence points which form diagonal hyper-surfaces; related to
    the predictability of the system
\\
    laminarity &
    $ LAM_{HS}=\frac{\sum_{v=v_{min}}^{N}vP(v)}{\sum_{v=1}^{N}vP(v)}$ &
    percentage of recurrence points which form vertical hyper-surfaces; related to
    the laminarity of the system
\\
    averaged diagonal hyper-surface size &
    $ L_{HS} = \frac{\sum_{l=l_{min}}^N l\, P(l)}{\sum_{l=l_{min}}^N P(l)}$ &
    related to the prediction length of the system
\\
    trapping size &	
    $TT_{HS} = \frac{\sum_{v=v_{min}}^{N} v P(v)} {\sum_{v=v_{min}}^{N} P(v)}$ &
    related to the size of the area in which the system does not change
\\
\hline
\end{tabularx}
\end{table}

\section{Extension to higher dimensions}
Now, we propose an extension of RPs to analyse higher
dimensional data. With this step we leave the RPs as a
method for investigating deterministic dynamics and
focus on its potential in
determining similar (recurrent) features in spatial data.

For a $d$-dimensional (Cartesian) system, we define
an $n$-dimensional recurrence plot by
\begin{equation}
\mathbf{R}_{\vec \imath,\vec \jmath} = \Theta\left(\varepsilon-\left\|\vec x_{\vec \imath} - \vec x_{\vec \jmath}\right\|\right),
\quad \vec x_{\vec \imath} \in \mathds{R}^m,\ \vec \imath,\,\vec \jmath \in \mathds{Z}^d,
\end{equation}
where $\vec \imath$ is the $d$-dimensional coordinate vector and $\vec x_{\vec \imath}$
is the phase-space vector at the location given by the coordinate vector $\vec \imath$.
This means that we decompose the spatial dimension of $\vec x_{\vec \imath}$
and consider each space direction separately, e.\,g.~$\vec x_{i_1, i_2, \ldots i_d}$ for
$i_1 = 1, \ldots, N$ but $i_2, \ldots, i_d = \text{const}$. Such vectors
are now one-dimensional curves in the $m$-dimensional space. 
Each of these vectors is pairwisely
compared with all others. These individual sub-RPs are the
components of the final higher-dimensional RP.
The resulting RP has now the dimension $n=2 \times d$ and cannot 
be visualised anymore. However, its quantification is still possible.

Similarly to the one-dimensional LOI given by $\mathbf{R}_{i,j} = 1 \ \forall \, i=j$,
we can find diagonally oriented, $d$-dimensional 
structures in this $n$-dimensional recurrence plot ($n=2\,d$), called
the {\it hyper-surface of identity} (HSOI):
\begin{equation}
\mathbf{R}_{\vec \imath,\vec \jmath} 
     \equiv  1 \quad \forall\  \vec \imath = \vec \jmath.
\end{equation}

In the special case of a two-dimensional image composed by scalar
values, we have 
\begin{equation}
\mathbf{R}_{i_1,i_2,j_1,j_2} 
     \equiv  \Theta\left(\varepsilon-\left\|x_{i_1,i_2} - x_{j_1,j_2}\right\|\right),
\end{equation}
which is in fact a four-dimensional recurrence plot, and its HSOI 
($\mathbf{R}_{i_1,i_2,i_1,i_2} \equiv 1$) is a two-dimensional plane.

\section{Quantification of Higher-Dimensional RPs}
The known RQA is based on the quantification
of the line structures in the two-dimensional RPs. Thus, the 
definition of higher-dimensional equivalent 
structures is crucial for a quantification analysis of
higher-dimensional RPs. 

Based on the definition of diagonal lines, Eq.~(\ref{eq_diagonalline}), 
we define a diagonal squared hyper-surface of size $\vec l$ ($\vec l = (l,\dots,l),
\vec l \in \mathds{Z}^d$) as
\begin{equation}\label{eq_diagonalline_x}
 \left( 1 - \mathbf{R}_{\vec \imath - \vec 1,\vec \jmath - \vec 1} \right) 
 \left( 1 - \mathbf{R}_{\vec \imath + \vec l,\vec \jmath + \vec l} \right) 
     \prod_{\substack{k_1, k_2,\dots,\\k_d=0}}^{l-1} \mathbf{R}_{\vec \imath+\vec k,\vec \jmath+\vec k}
     \equiv 1.
\end{equation}
In particular, for the two-dimensional case such a diagonal 
hyper-surface (HS) is thus defined as
\begin{equation}
 \left( 1 - \mathbf{R}_{i_1-1, i_2-1, j_1-1, j_2-1} \right) 
 \left( 1 - \mathbf{R}_{i_1+l, i_2+l, j_1+l, j_2+l} \right) 
     \prod_{k_1, k_2=0}^{l-1} \mathbf{R}_{i_1+k_1, i_2+k_2, j_1+k_1, j_2+k_2}
     \equiv 1.
\end{equation}

The next characteristic structure, the vertical squared HS
of size $\vec v$ ($\vec v = (v,\dots,v), \vec v \in \mathds{Z}^d$)
is defined as
\begin{equation}\label{eq_verticalline_x}
 \left( 1 - \mathbf{R}_{\vec \imath,\vec \jmath - \vec v} \right) 
 \left( 1 - \mathbf{R}_{\vec \imath,\vec \jmath + \vec v} \right) 
     \prod_{\substack{k_1, k_2,\dots,\\k_d=0}}^{v-1} \mathbf{R}_{\vec \imath,\vec \jmath+\vec k}
     \equiv 1.
\end{equation}
Its 2D equivalent is
\begin{equation}
 \left( 1 - \mathbf{R}_{i_1, i_2, j_1-1, j_2-1} \right) 
 \left( 1 - \mathbf{R}_{i_1, i_2, j_1+v, j_2+v} \right) 
     \prod_{k_1, k_2=0}^{v-1} \mathbf{R}_{i_1, i_2, j_1+k_1, j_2+k_2}
     \equiv 1.
\end{equation}

Using these definitions, we can construct the frequency distributions $P(l)$ and $P(v)$
of the sizes of diagonal and vertical HS in the 
higher-dimensional RP. This way we get generalised RQA measures $DET_{HS}$, $LAM_{HS}$,
$L_{HS}$ and $TT_{HS}$ as defined in Tab.~\ref{eq_rqa},
which are now suitable for characterising spatial (e.\,g.~two-dimensional) data.

\begin{figure}[bp] 
\centering \includegraphics[width=\columnwidth]{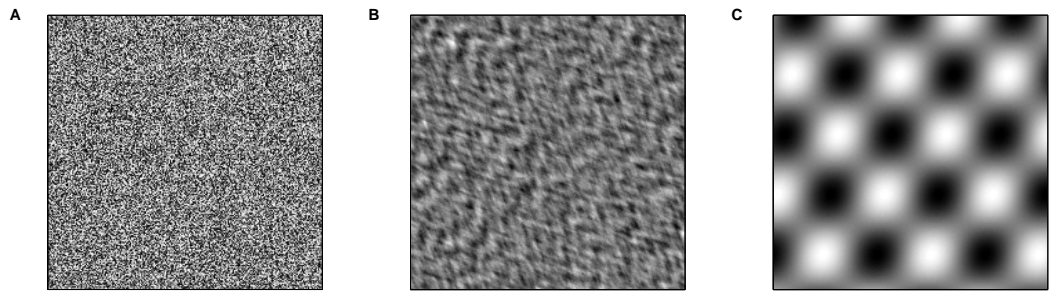} 
\caption{Two-dimensional prototypical examples: test 
images representing (A) uniformly distributed 
white noise, (B) a two-dimensional auto-correlated process (2D-AR2) and (C) 
periodical recurrent structures.}\label{fig_data_sample}
\end{figure}

\section{Model Examples}

In order to illustrate the ability of the proposed high-dimensional
RP's extension, we consider
three prototypical model examples from 2D image analysis. The first image (A)
is produced by uniformly distributed white noise, the second one (B) is 
the result of a two-dimensional auto-correlated process of 2nd order 
(2D-AR2; $x_{i, j} = \sum_{k,l=1}^2 a_{k,l}\,x_{i-k,j-l} + \xi$, where $a_{k,l}$ is
the 2D matrix of model parameters and $\xi$ is Gaussian white noise) 
and the third one (C) represents periodical recurrent 
structures (Fig.~\ref{fig_data_sample}). All these example images have a 
size of $200 \times 200$~pixels and 
are normalised to a mean of zero and a standard
deviation of one.

The resulting RPs are four-dimensional matrices of size 
$200 \times 200 \times 200 \times 200$, and can hardly be visualised.
However, in order to visualise these RPs, we can reduce their dimension by one
by considering only those part of the RPs, where 
$i_2 = j_2$. The resulting $200 \times 200 \times 200$ cube is a 
hypersurface of the four-dimensional RP along the LOI. 
For the threshold we use $\varepsilon = 0.2$, which
gives clear representations of the RPs.

\begin{figure}[tbp] 
\centering 
\scriptsize{\textsf{A}}\includegraphics[width=0.31\columnwidth]{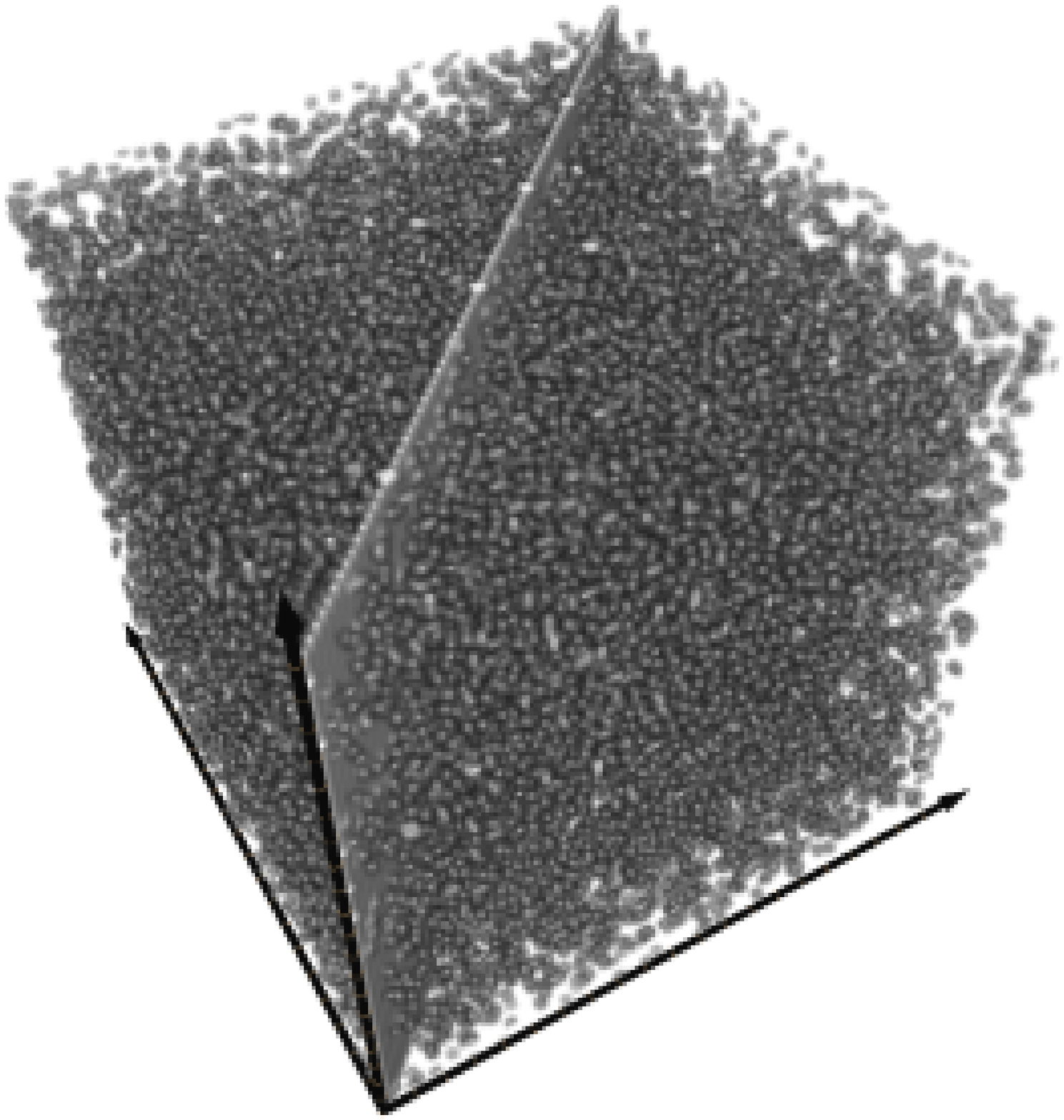} 
\scriptsize{\textsf{B}}\includegraphics[width=0.31\columnwidth]{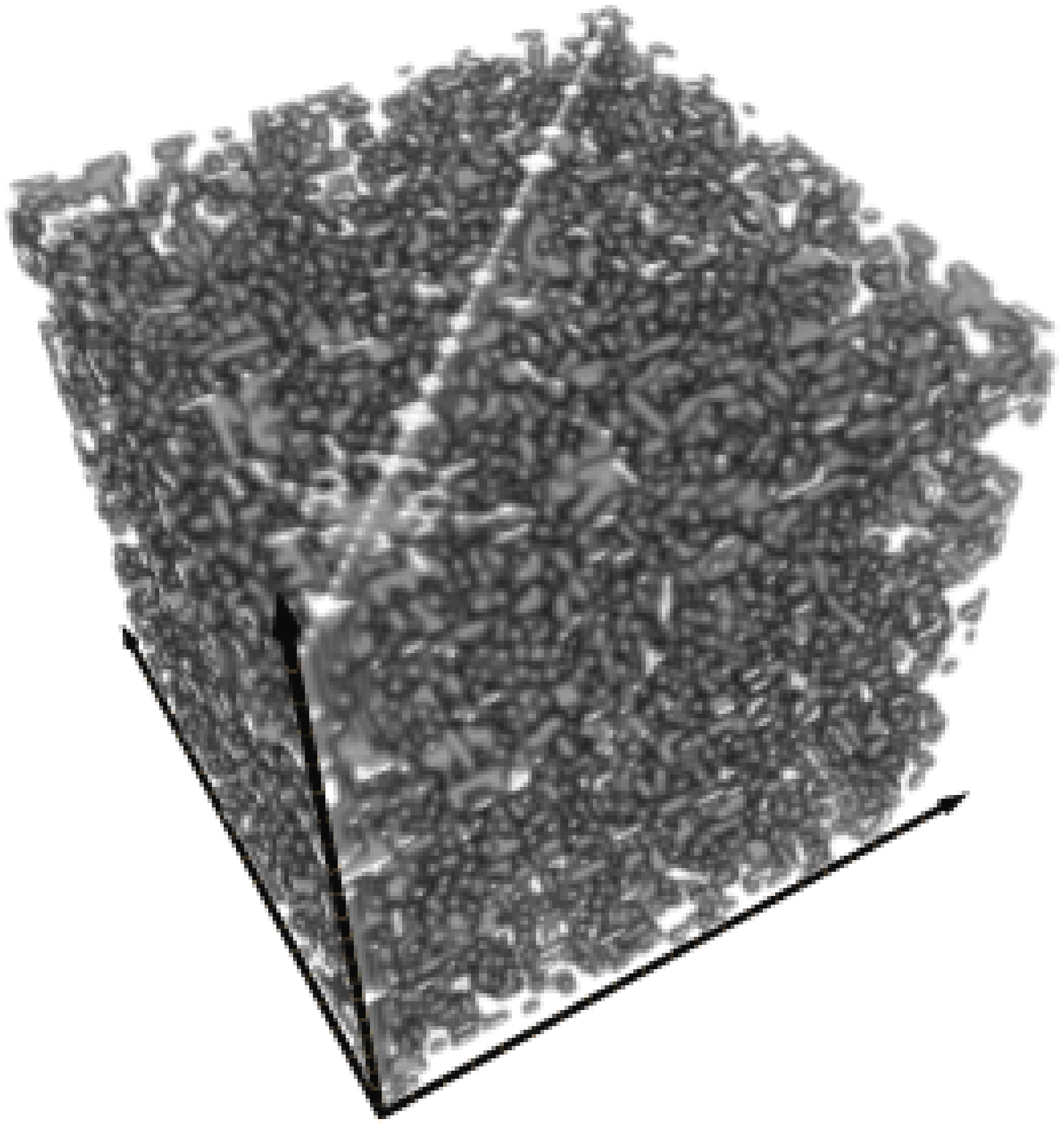} 
\scriptsize{\textsf{C}}\includegraphics[width=0.31\columnwidth]{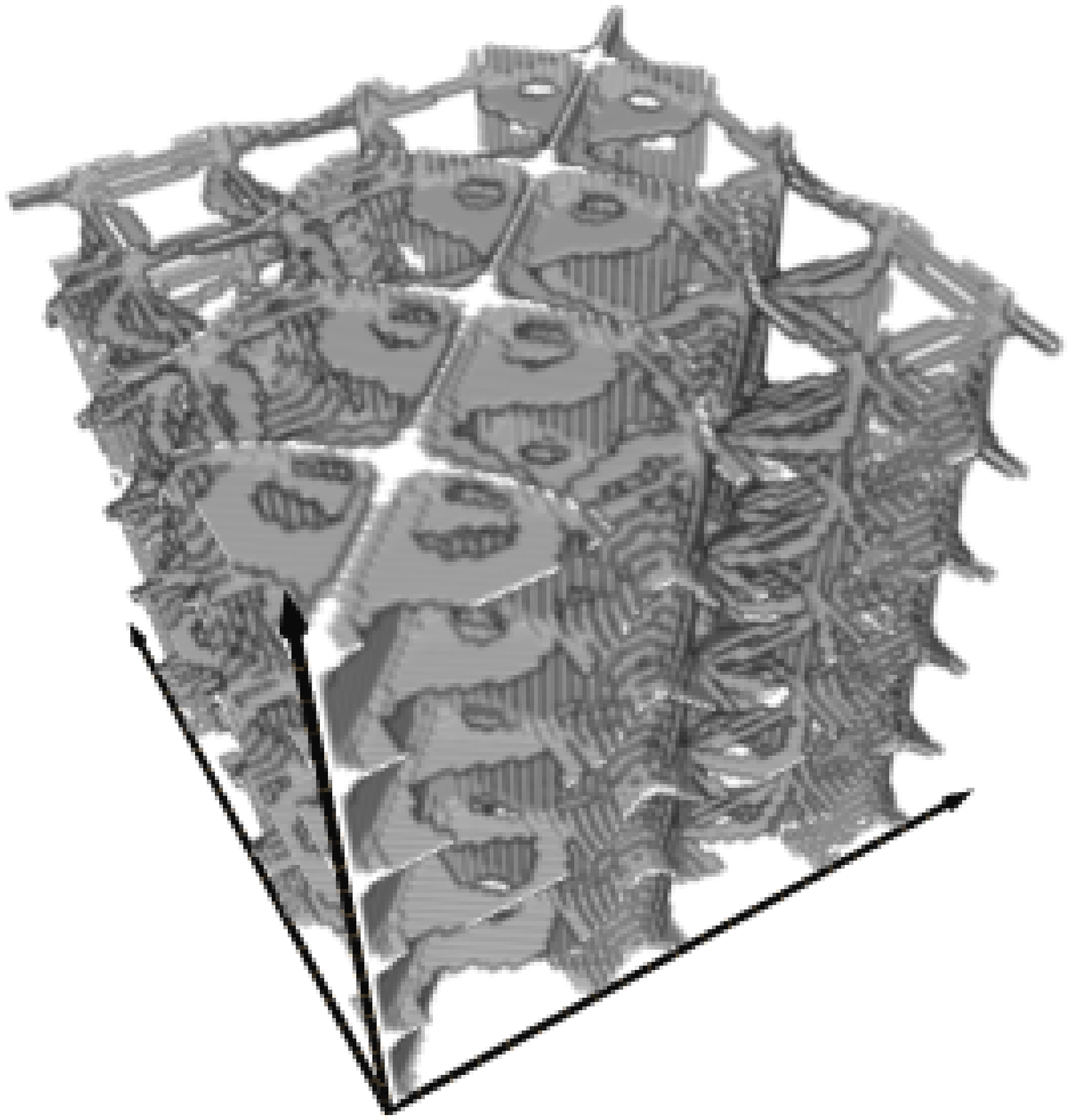} 
\caption{Three-dimensional subsections of four-dimensional RPs of the 
images shown in Fig.~\ref{fig_data_sample}. As known from
one-dimensional RQA, (A) random data causes homogeneous RPs consisting
of single, dis-connected points, (B) correlations in data 
cause extended connected structures and
(C) periodic data induce periodically occuring structures in the 
RPs.}\label{fig_rp_sample}
\end{figure}

\begin{figure}[btp] 
\centering 
\scriptsize{\textsf{A}}\includegraphics[width=0.31\columnwidth]{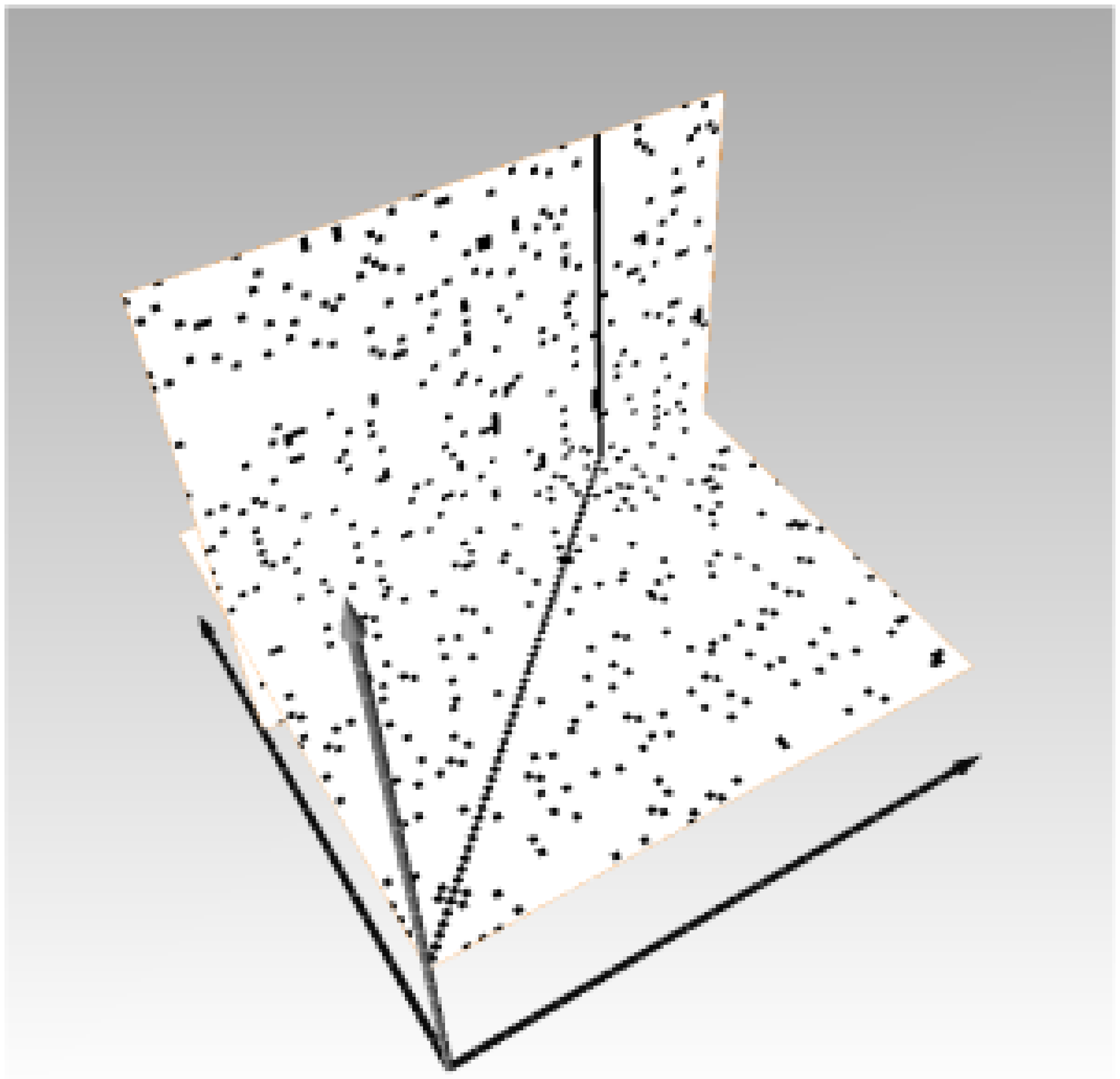} 
\scriptsize{\textsf{B}}\includegraphics[width=0.31\columnwidth]{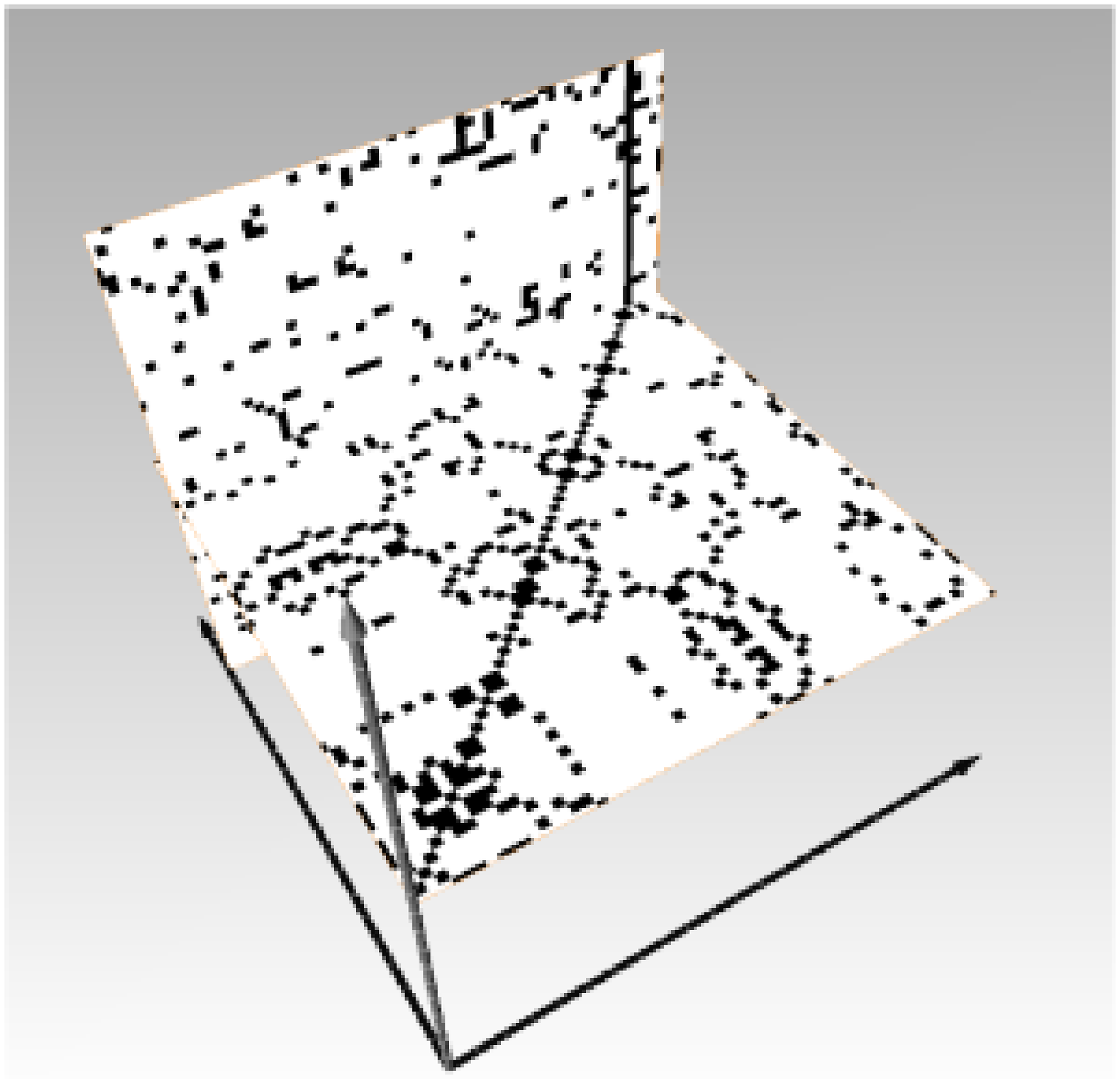} 
\scriptsize{\textsf{C}}\includegraphics[width=0.31\columnwidth]{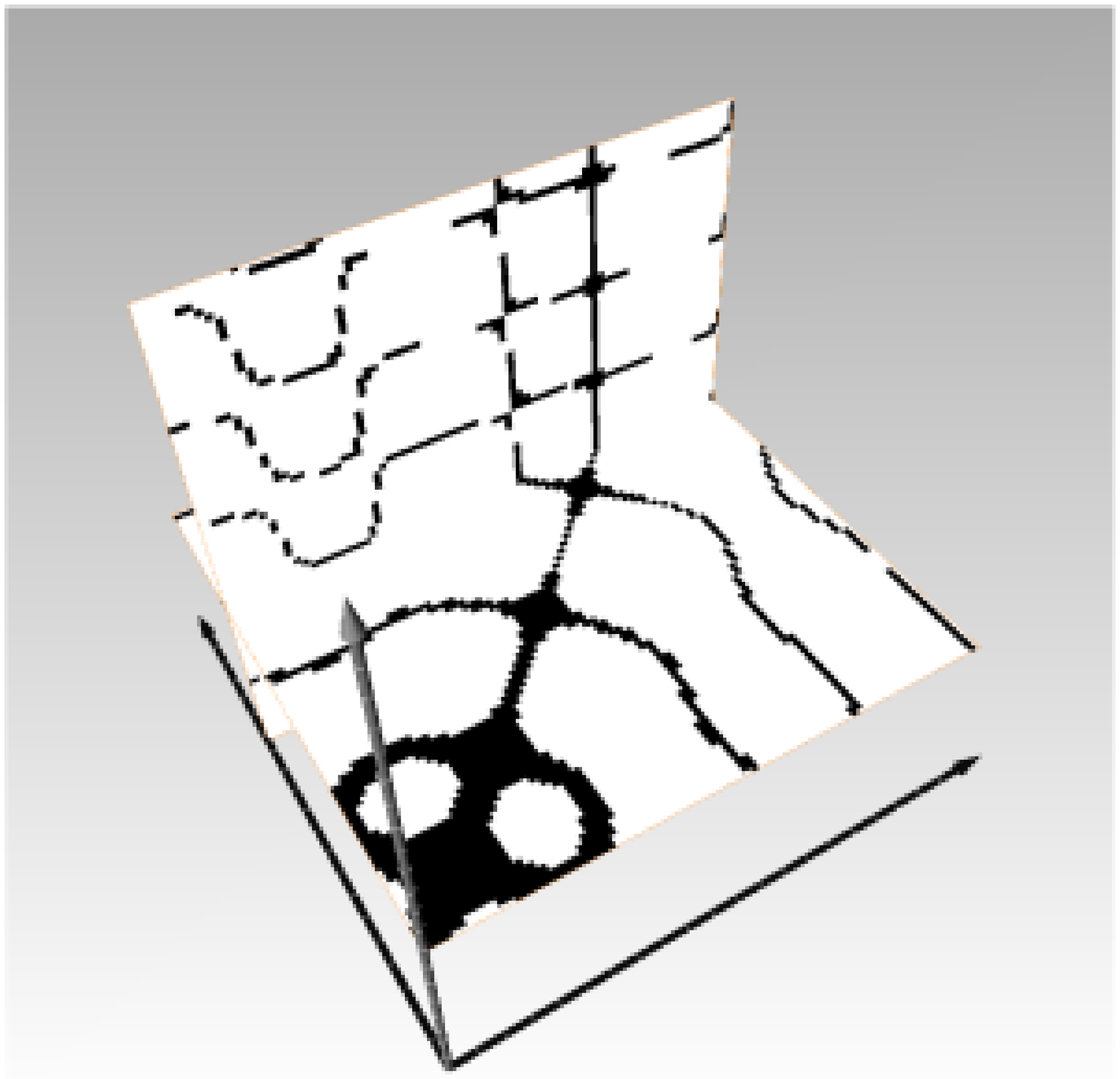} 
\caption{Slices of the subsections of the four-dimensional RPs 
shown in Fig.~\ref{fig_rp_sample}. The similarity to known recurrence 
plots is obvious: (A) noise, (B) auto-correlated data and (C) periodic data.}\label{fig_rp2_sample}
\end{figure}

The features occuring in higher-dimensional RPs provide similar
information as known from the classic one-dimensional RPs. 
Separated single points correspond
to strongly fluctuating, uncorrelated data as it is typical 
for, e.\,g., white noise (Fig.~\ref{fig_rp_sample}A).
Auto-correlations in data cause extended structures, which can be
lines, planes or even cuboids (Fig.~\ref{fig_rp_sample}B).
Periodical recurrent patterns in data imply periodic line and
plane structures in the RP (Fig.~\ref{fig_rp_sample}C). Two-dimensional
slices through such RPs contain similar patterns found by common
RPs (Fig.~\ref{fig_rp2_sample}).

We compute the proposed RQA measures (Tab.~\ref{eq_rqa})
for the histograms of the sizes of diagonal and vertical 
planes (2D HS) in the four-dimensional RPs. For all three examples
we use for the minimal size of the diagonal and 
vertical HS $l_{min}=3$\,pixels and $v_{min}=4$\,pixels. Although the RQA measures
depend on the value of $\varepsilon$, its selection is not crucial for
our purpose to discriminate the three different types of structures
in the test images. The chosen values
for $l_{min}$ and $v_{min}$ are found to be optimal for discriminating
the considered images. By choosing smaller values of 
$l_{min}$ and $v_{min}$ (but larger than one), the 
measures $DET_{HS}$ and $LAM_{HS}$ are closer for 
the 2D-AR2 and the periodic image.

\begin{table}
\caption{Recurrence quantification measures for the prototypical 
examples shown in Fig.~\ref{fig_data_sample}. The measures 
are explained in Tab.~\ref{eq_rqa}.}\label{tab_res_examples}
\centering \begin{tabular}{lrrrrr}
\hline
Example&	$RR$&	$DET_{HS}$&	$LAM_{HS}$&	$L_{HS}$&	$TT_{HS}$\\
\hline
\hline
(A) noise	&0.218&	0.007&	0.006&	3.7&	3.0\\
(B)	2D-AR2	&0.221&	0.032&	0.065&	3.1&	3.1\\
(C)	periodic&0.219&	0.322&	0.312&	5.8&	5.6\\
\hline
\end{tabular}
\end{table}

Four of five RQA
measures clearly discriminate between the three types of images 
(Tab.~\ref{tab_res_examples}).
Only the recurrence rate $RR$ is roughly the same for all test objects. This is
because all images were normalised to the same standard deviation.
For the random image (A) the determinism $DET_{HS}$ and
laminarity $LAM_{HS}$ tend to zero, what is expected, because the values
in the image heavily fluctuate even between adjecent pixels. 
For the 2D-AR2 image (B), $DET_{HS}$ and $LAM_{HS}$ are slightly above zero, revealing
the correlation between adjecent pixels. The last example (C)
has, as expected, the highest values in $DET_{HS}$ and $LAM_{HS}$, because
same structures occur many times in this image and the image is 
rather smooth. Although the trend
in $DET_{HS}$ and $LAM_{HS}$ is similar, there is a significant difference
between both measures. Whereas $LAM_{HS}$ represents the probability that a
specific value will not change over spatial variation (what results
in extended same-coloured areas in the image), $DET_{HS}$ measures 
the probability that similar changes in the image recur. $LAM_{HS}$ is twice
of $DET_{HS}$ for the 2D-AR2 image, obtaining that there are more areas
without changes in the image than such with typical, recurrent changes.

\section{Application to pQCT data of proximal tibia}

According to the definition of the World Health Organisation, 
osteoporosis is a disease characterised by bone loss and changes in the
structure of the bone. In the last years, the focus
changed to structural assessment of the trabecular bone, because
bone densitometry alone is very limited to explain all variation in bone 
strength. Furthermore, the rapid progress in the development of new
high-resolution computer tomography (CT) scanners 
facilitates investigations of the bone micro-architecture. 
Different approaches using methods coming from
nonlinear dynamics have been recently proposed in order to
evaluate structural changes 
\cite{benhamou1994,saparin1998,gowin1998,dougherty2001} or 
even to predict fracture risks or biomechanical properties
\cite{haire1998,majumdar1999,prouteau2004}. These approaches
use, e.\,g., scaling properties of bone micro-structure
or symbol-encoding of the bone architecture.

Using the RP based method, we will focus here on the  
recurrent structures found in images of trabecular bone
of proximal tibiae obtained by peripheral 
quantitative computer tomography (pQCT). The images were
aquired from bone specimens with different stages of 
osteoporosis as assessed by 
bone mineral density (BMD). Being applied to such images, the RP provides 
information about recurrences of bone and soft tissue.

\begin{figure}[hbtp] 
\centering 
\includegraphics[width=0.4\columnwidth, angle=180]{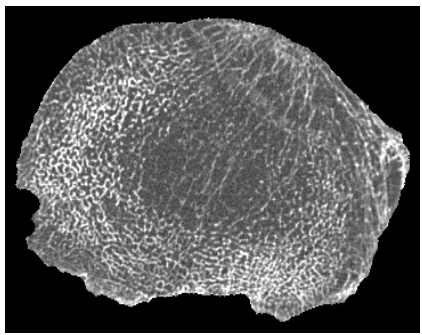} 
\caption{Typical axial pQCT slice of human proximal tibia acquired 17~mm 
below the tibial plateau. The trabecular BMD
is 65.5\,mg/cm$^3$.}\label{fig_tibia}
\end{figure}

The spatial recurrence analysis is applied to high-resolution
pQCT axial slices of human proximal tibia acquired 17~mm below the
tibial plateau, with pixel size 
200\,$\mu$m and slice thickness 1\,mm (Fig.~\ref{fig_tibia}). 
The images were acquired from 25 bone specimens 
with a pQCT scanner XCT-2000 (Stratec GmbH, Germany).
The trabecular bone mineral density of these specimens ranges from
30 to 150\,mg/cm$^3$.
A standardised image pre-processing procedure was applied to exclude the 
cortical shell from the analysis \cite{saparin1998,saparin2002}. The RQA
was computed by using the parameters $\varepsilon = 0.04$\,cm$^{-1}$, 
$l_{min} = v_{min} = 400$\,$\mu$m. These minimal lengths correspond
to two pixels and is found to be appropriate for pQCT images of 
such resolution.

In order to further evaluate the proposed RQA measures, we compare them
with some recently 
introduced structural measures of complexity (SMCs) \cite{saparin1998,saparin2002}. 
The SMCs are based on a symbol-encoding of bone elements in
the pQCT image. Here we focus on the following SMCs:
\begin{enumerate}
\item Entropy ($S_a$): quantifies the probability distribution of X-ray attenuation within the ROI;
\item Structure Complexity Index (SCI): assesses the complexity and homogeneity of the
structure as a whole;
\item Trabecular Network Index (TNI) evaluates richness, orderliness, and homogeneity of
the trabecular network.
\end{enumerate}
The computation of the SMCs is applied on the same trabecular
area like the RQA measures.

\begin{figure}[hbtp] 
\centering 
\includegraphics[width=.5\columnwidth]{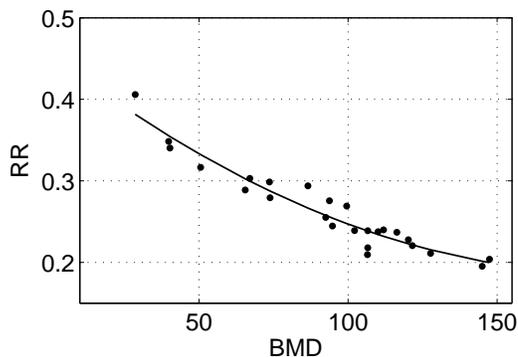} 
\caption{Recurrence rate $RR$ obtained from
four-dimensional RPs of pQCT images of trabecular bone in
human proximal tibia of different osteoporotic stages.}\label{fig_rr_bmd}
\end{figure}

\begin{figure}[hbtp] 
\centering 
\includegraphics[width=\columnwidth]{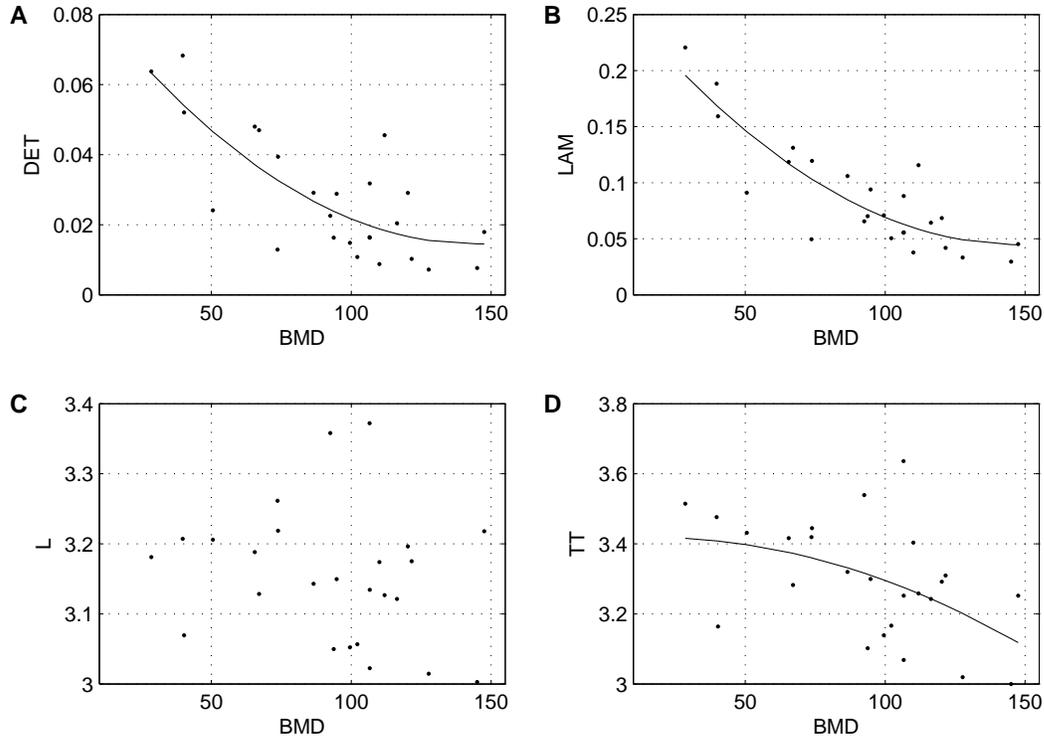} 
\caption{Determinism $DET$ (A), laminarity $LAM$ (B),
mean line length $L$ (C) and trapping time $TT$ (D) obtained from
four-dimensional RPs constructed from pQCT images of trabecular 
bone in human proximal tibiae
with different degree of osteoporosis.}\label{fig_rqa_bmd}
\end{figure}

\begin{figure}[hbtp] 
\centering 
\includegraphics[width=\columnwidth]{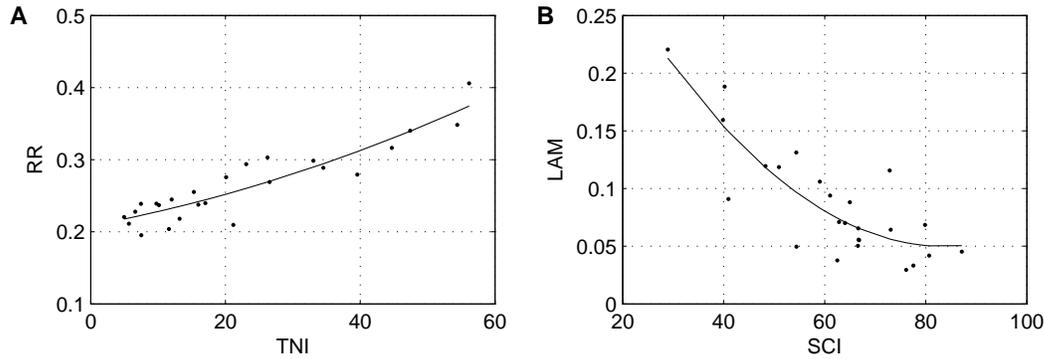} 
\caption{Recurrence rate $RR$ (A) and laminarity $LAM$ (B)
compared with trabecular network index TNI and
structure complexity index (SCI).}\label{fig_rqa_smc}
\end{figure}

The application of the recurrence plot extension to the pQCT images
of proximal tibiae reveals a relationship between the recurrences
in the trabecular architecture and the osteoporotic stage
(Fig.~\ref{fig_rr_bmd} and Tab.~\ref{tab_correlation}). 
$RR$ is largest for osteoporotic bone and 
shows the strongest relationship
with the degree of osteoporosis: it is clearly anti-correlated with BMD (Spearman's 
rank order correlation coefficient $R=-0.94$). $DET_{HS}$ and $LAM_{HS}$ are 
also maximal for tibiae with high degree of osteoporosis 
($R=-0.66$ and $-0.79$; Fig.~\ref{fig_rqa_bmd}). 
We do not find a strong relation between $L_{HS}$, $TT_{HS}$ and BMD.
The comparison with the SMCs reveals good relationships 
between the RQA measures and $S_a$, SCI and TNI 
(Fig.~\ref{fig_rqa_smc} and Tab.~\ref{tab_correlation}).
Thus, the RQA measures $RR$, $DET_{HS}$ and $LAM_{HS}$ 
contain also information about the
complexity and homogeneity of the trabecular network.

Thus, the proposed RP approach reveals that during the development
of osteoporosis the
structures in the corresponding pQCT image become more and
more recurrent. This is in a good agreement with a decreasing 
complexity in the micro-architecture of bone. It confirms
the results of an analysis of pQCT images acquired from human proximal
tibia and lumbar vertebrae based on symbolic dynamics 
\cite{saparin1998,saparin2002}. The direct comparison
with the structural quantities (SMCs) shows that the RQA measures
provide information about the bone architecture. 
The RQA measures reveal a low rate of change 
for bone of higher BMD, but higher rate of changes 
for specimens with lower BMD (Figs.~\ref{fig_rr_bmd} and \ref{fig_rqa_bmd}).
This reflects a higher sensitivity of these measures for
osteoporotic trabecular bone and emphasises the nonlinear
relationship between the bone architecture as assessed by the
RQA measures and bone mass as evaluated by the BMD. As it has been
recently shown that the SMCs provide a better estimation of the
mechanical bone strength than BMD alone \cite{saparin2005a}, 
the proposed RQA measures could further enhance the evaluation to
assess the fracture risk of bone.

\begin{table}
\caption{Rank correlation coefficients $R$ for recurrence
quantification measures, BMD and
structural measures of complexity (only significant values
are shown).}\label{tab_correlation}
\centering \begin{tabular}{lrrrrrrrr}
\hline
2D-RQA	&BMD		&$S_a$	&SCI	&TNI\\
\hline
\hline
RR		&$-0.94$	&$-0.92$	&$-0.91$&0.84\\
DET		&$-0.65$	&$-0.58$	&$-0.61$&0.61\\
LAM		&$-0.78$	&$-0.73$	&$-0.75$&0.72\\
L		&--			&--			&--		&--	\\
TT		&$-0.57$	&$-0.51$	&--		&0.49\\

\hline
\end{tabular}
\end{table}

\subsection{Conclusions}

A generalisation of the method of recurrence plots (RPs) and
recurrence quantification analysis (RQA) for 
the application to higher-dimensional spatial data has been proposed here. 
This new method can be used for 2D image analysis,
in particular to reveal and quantify 
recurrent structures in 2D images.
Applying this method on model images, we have shown that it 
is able to distinguish typical spatial structures by means of recurrences. 
As a first application, we have used this method for the
comparison of CT images of human proximal tibia with
different degree of osteoporosis. We have found a clear
relationship between some of the proposed RQA measures
and the complexity and homogeneity of the trabecular structure.
Moreover, this approach can be easily extended to higher 
dimensions, e.\,g., for 3D analysis of micro-CT
images of human bone.
This approach will be the base for the further development of 
methods for the assessment of structural alteration in trabecular 
bone with osteoporosis in patients on Earth
or in space flying personnel in microgravity conditions.

\section{Acknowledgments}
This study was supported by grants from project MAP AO-99-030 
(contract \#14592) of the Microgravity 
Application Program/Biotechnology from the Human Spaceflight Program 
of the European Space Agency (ESA) and by the European Union
through the Network of Excellence BioSim, contract LSHB-CT-2004-005137\&\#65533. 
The authors would also like 
to acknowledge Scanco Medical AG, Siemens AG, and Roche Pharmaceuticals 
for support of the study and thank Wolfgang Gowin and Erika May for preparation
and scanning of the bone specimens.

\clearpage

\bibliographystyle{elsart-num}
\bibliography{mybibs,rp,bone}

\end{document}